# INFLUENCE OF VACANCIES ON THE NANOPARTICLE PULSATION

M.A.Korshunov

*Institute of physics it. L. V.Kirenskogo of the Siberian Branch of the Russian Academy of Science*
*Krasnoyarsk, Academgorodok, 660036, Russia*
*e-mail:mkor@iph.krasn.ru*

Computer modeling of the one-dimensional and three-dimensional nanoparticles with Van-der-Waals interaction was performed. The arrangement of atoms was defined on the grounds of an energy minimum. The calculations have shown that in the presence of vacancies in a nanoparticle and the account of a relaxation and oscillations the nanoparticle will pulse. This pulsation on distance between atoms of a nanoparticle is observed in both one-dimensional and a three-dimensional cases.

Our calculations have shown that the distance between the nearest atoms depends on the number of atoms. In Fig. 1, the distance change Δa between the nearest atoms at centre of the linear chain with increase of number of atoms surrounding them is shown. Interaction between atoms was considered not only between the nearest neighbors, and the energy minimization on distance between them was done. This curve depends on the potential shape. Such behavior of parameters of a lattice agrees with the experimental data [1]. For a three-dimensional case, similar effects are obtained. In the considered nanostructures with the reduction of their sizes, the increase of the disorder of atoms positions is observed. With the increase of the sizes of nanostructures the lattice becomes more ordered.

Here the question arise: if there are a vacancies in the nanocrystal, how the lattice parameters will change?

To answer this question, we have made a computer modeling of the one-dimensional and three-dimensional nanoparticles with Van-der-Waals interaction and with the presence of vacancies in the structure. Interaction between atoms was taken into account within the method of atom-atom potentials.

First, we considered the diffusion of atom towards the vacancy without taking into account the relaxation of atoms. Resulting potential energy ΔU dependence on distance Δr is shown in Fig. 2.

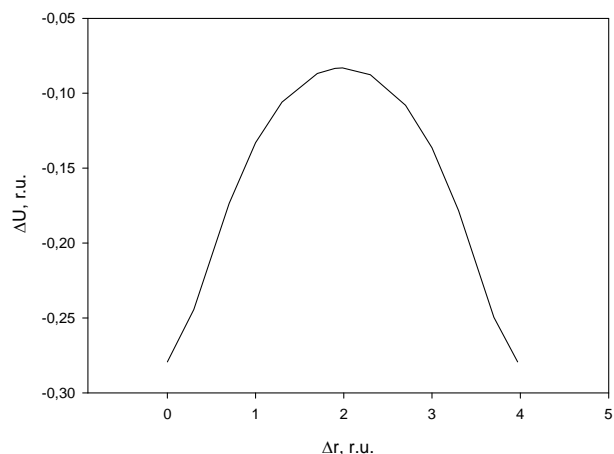

**Fig. 2.** Change of a potential energy of an atom at the transition to the next vacancy.

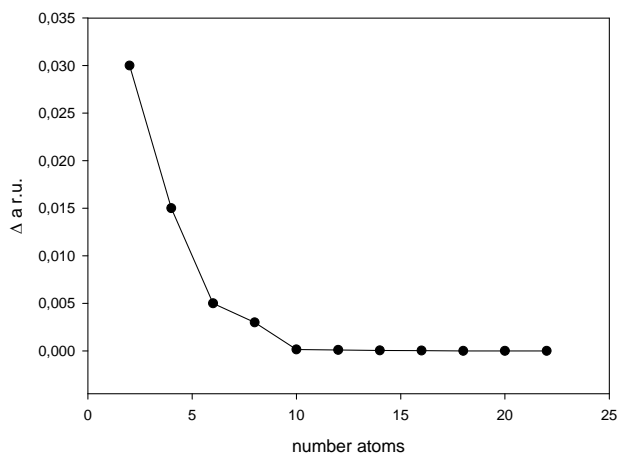

**Fig. 1.** The change of the distance between the nearest atoms, Δa, at centre of the linear chain, depending on number of atoms is shown.

Second, we considered thermal oscillations of atoms, due to which the atom can move from a steady position towards a vacancy, and a relaxation of surrounding atoms. Resulting ΔU(Δr) dependence is presented in Fig. 3. Note, in this situation we have a potential well, not a potential barrier.

The exit from the well is possible due to thermal oscillations of atoms.

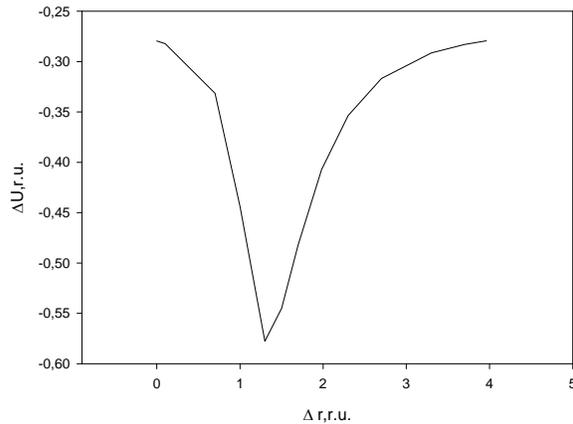

**Fig. 3.** Change of a potential energy of an atom at transition in the next vacancy with the account of a relaxation of atoms.

As the next step, it is necessary to consider disorder in an arrangement of atoms in a nanoparticle with the reduction of nanoparticle's size. Besides the potential energy change, in this case there is a distance change between particles that is a consequence of a relaxation. In Fig. 4 the distance change between atom and its nearest neighbors, ΔR, is shown for the case when the atom goes to the next vacancy.

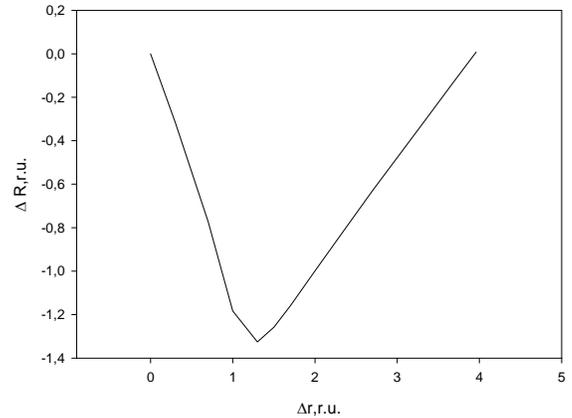

**Fig. 4.** Distance change between atom and its nearest neighbours at atom transition in the next vacancy with the account of a relaxation of atoms.

If there are a small number of atoms in a nanoparticle, then after the shift of the nearest neighbors there is a shift of other atoms up to a nanoparticle boundary. Therefore, there is a periodic squeezing and expansion of a nanoparticle (nanopulsar). Apparently, this should influence other physical properties of the nanoparticle.

Summarizing, we have demonstrated that if there are vacancies in a nanoparticle then there can be a periodic squeezing and expansion of the nanoparticle.